\documentclass[preprint2]{aastex}

\def\msol{M_\odot}
\def\mjup{M_{\rm J}}
\def\amax{\alpha_{\rm max}}
\def\mc{M_{\rm c}}
\def\mi{M_{\rm i}}
\def\mf{M_{\rm f}}
\def\etot{E_{\rm tot}}
\def\edeut{E_{\rm deut}}
\def\mdot{\dot M}
\def\tmdot{\tau_{\rm \dot M}}
\def\tkh{\tau_{\rm KH}}
\def\msolyr{M_\odot \rm {yr}^{-1}}
\def\msun{M_\odot}
\def\rsun{M_\odot}
\def\rsun{R_\odot}

\def\tef{T_{\mathrm{eff}}}
\def\te{T_{\mathrm{eff}}}
\def\mdot{\dot{M}}
\def\dt{\Delta t}
\def\minit{m_{\rm i}}
\def\be{\begin{equation}}
\def\ee{\end{equation}}
\def\mjup{M_\mathrm{{Jup}}}

\def\lsun{L_\odot}

\def\simgr{\,\hbox{\hbox{$ > $}\kern -0.8em \lower 1.0ex\hbox{$\sim$}}\,}
\def\simle{\,\hbox{\hbox{$ < $}\kern -0.8em \lower 1.0ex\hbox{$\sim$}}\,}

\shorttitle{Episodic accretion}
\shortauthors{Baraffe et al.}

\begin{document}

\title{ Observed luminosity spread in young clusters and Fu Ori stars: a unified picture}

\author{I. Baraffe\altaffilmark{1}, E. Vorobyov\altaffilmark{2,}\altaffilmark{3} and
G. Chabrier\altaffilmark{4,1}}
\altaffiltext{1}{University of Exeter, Astrophysics group, EX4 4QL Exeter; i.baraffe@ex.ac.uk}
\altaffiltext{2}{Institute of Astrophysics, The University of Vienna, Vienna, 1180, Austria; 
eduard.vorobiev@univie.ac.at.}
\altaffiltext{3}{Research Institute of Physics, Southern Federal University, Rostov-on-Don, Russia}
\altaffiltext{4}{\'Ecole Normale Sup\'erieure, Lyon, CRAL (UMR CNRS 5574), Universit\'e de Lyon, France;
gilles.chabrier@ens-lyon.fr}


\begin{abstract}
The idea that non steady accretion during the embedded phase of protostar evolution can produce 
the observed luminosity spread in the Herzsprung-Russell diagram (HRD) of young clusters has recently been 
called into question. 
Observations of Fu Ori, for instance, suggest an expansion of the star 
during strong accretion events whereas the luminosity spread implies a contraction of the accreting objects, decreasing their radiating surface.  In this paper, we present a global scenario based on 
calculations coupling episodic accretion histories derived from 
numerical simulations of collapsing cloud prestellar cores of various masses
and subsequent protostar evolution. 

Our calculations show that, assuming an initial protostar mass $\mi \sim 1\,\mjup$, typical of the second Larson's core, both the luminosity spread in the HRD and the inferred properties of Fu Ori events (mass, radius, accretion rate) can be explained by this scenario, providing two conditions. First, there must be some
 variation within the fraction of  accretion energy absorbed by the protostar during the accretion process. Second the range of this variation should increase with increasing accretion burst intensity, and thus with the initial core mass and final star mass. The numerical hydrodynamics simulations of collapsing cloud prestellar cores indeed show that the intensity of the accretion bursts correlates with the mass and  initial angular momentum
of the  prestellar core. Massive prestellar cores with high initial angular momentum are found to produce intense bursts characteristic of Fu Ori like events. Our results thus suggest a link between  the burst intensities and the fraction of accretion energy absorbed by the protostar, with some threshold in the accretion rate, of the order of $10^{-5}\msolyr$, delimitating the transition from "cold" to "hot" accretion. Such a transition might reflect a change in the accretion geometry with increasing accretion rate, i.e. a transition from magnetospheric or thin disk to thick disk accretion, or in the magnetospheric interaction between the star and the disk. Conversely, the luminosity spread can also be explained by a variation of the initial protostar mass within the $\sim 1$-5$\,\mjup$ range, although it is unclear for now whether such a spread among second Larson's core can be produced during the prestellar core second collapse.
This unified picture confirms the idea that early accretion during protostar and proto-brown dwarf formation/evolution can explain the observed luminosity spread in young clusters without invoking any significant age spread, and that the concept of a well defined birthline does not apply for low-mass objects.
Finally, we examine the impact of accretion on the determination of the IMF in young clusters.
\end{abstract}


\section{Introduction} \label{intro}

Recently, Baraffe, Chabrier \& Gallardo (2009, BCG09) suggested
that early phases of accretion during the embedded phases of protostellar evolution could explain the  luminosity spread observed in young, a few Myr old, clusters and star forming regions. The same scenario also provides a natural explanation for the unexpected lithium depletion observed in some young objects (Baraffe \& Chabrier 2010, BC10). 
Given the increasing evidence that low-mass protostars accrete a substantial part of their material through short intense bursts of accretion (Kenyon et al. 1990; Enoch et al. 2009; Evans et al. 2009), due to gravitational  (Vorobyov \& Basu 2005, 2006; Machida et al. 2010) or a combination of gravitational and magneto-rotational (Zhu et al. 2009) instabilities in the accretion disk,
the scenario proposed by BCG09 is based on the process of so-called "episodic accretion". Assuming that the accreting material radiates away essentially all its internal energy at the shock front, with no contribution to the internal heat content of the accreting object (the so-called "cold accretion" process, see \S 3), and taking for the initial protostar a mass range from 1 $\mjup$ to 0.1 $\msun$, BCG09 were able to produce a spread in luminosity in the Herzsprung-Russell diagram (HRD) for the final objects, at the end of the accretion process, i.e. after a few Myr, equivalent to a typical $\sim$ 10 Myr age spread for non-accreting objects. Their major conclusion was thus that the observed HRD luminosity spread does not stem from an age spread, but rather from the impact of early accretion history on the structure and the cooling of the protostar. This conclusion is supported by the recent observational analysis of stars with and without disks in the Orion Nebula cluster by Jeffries et al. (2011).  Finding no significant difference in the mean ages or age distributions inferred for stars with and without disks, respectively, these authors conclude that the observed luminosity dispersion cannot be due to a large age spread, but rather stems from a combination of observational uncertainties and physical mechanisms. 

Conducting a study similar to BCG09, Hosokawa et al. (2011, HOK11) recently reexamined the impact of early accretion on the evolution of low mass stars (LMS) and brown dwarfs (BD). Although these authors find results similar to BCG09 concerning the effects of accretion on the structure of these objects, confirming BCG09's study, they reach different conclusions and claim that accretion cannot produce the observed spread in the coolest part of the HRD, {\it i.e.,} for objects with $\te \simle$ 3500 K. 
Further contributing to the debate, Hartmann et al. (2011, HZC11) recently reanalysed the spectral energy distribution of FU Orionis. This object undergoes short accretion bursts from a disk and provides an excellent laboratory to test the effects of strong accretion bursts on the structure of accreting objects. HZC11 highlight the absence of a hot boundary layer emission, which is best explained by significant heating of the protostar upper layers by the disk infalling material.  Furthermore, HZC11 infer a rather large radius ($\sim 5$ R$_\odot$) for the protostar whereas, in case of aforementioned cold accretion, BCG09 predict a much more compact structure, because of the accretion-induced contraction of the object. HZC11 thus conclude that for accretion histories with burst rates as high as about $10^{-4} \msolyr$, as observed in Fu Ori objects, some fraction of the accreting energy must be absorbed by the protostar (departing from the genuine "cold accretion" regime), yielding an expansion of the object, as indeed obtained by BCG09's in the "hot-accretion" case.

In order to clarify these controversies, we present in this paper a global scenario, based on the 
coupling between episodic accretion calculations obtained from collapsing cloud prestellar  core simulations, and  accreting protostar/BD evolutionary models. We show that such a scenario explains both the HRD luminosity spread {\it and} the expansion of Fu Ori stars, reinforcing the conclusions reached in BCG09. We also explain why HOK11 reached different conclusions than BCG09, even though conducting similar calculations, which should help resolving the controversy raised about this issue.

\section{Accretion histories from prestellar core collapse calculations}\label {accretion}

In the present calculations, the accretion rates on the protostar, $\dot{M}(t)$, are obtained from 
hydrodynamics simulations of {\it prestellar core} collapse  similar to the ones conducted 
in Vorobyov \& Basu (2010, VB10) and Vorobyov (2010, V10). 
All the details of the model can be found in VB10. The simulations start from collapsing prestellar
cores of mass $\mc$ with Bonnor-Ebert like density profiles, and follow the formation and 
evolution of the protostar (mimicked as a point mass gravitating object at the centre 
of a sink cell of radius $r_{\rm sc}=6$~AU\footnote{Our test runs have shown that decreasing $r_{\rm sc}$ 
by a factor of three does not qualitatively change the accretion pattern but entails
almost a factor 10 increase in computational time, making simulations of a large number
of models with small
sink cells prohibitively expensive.} ), the disk and the envelope. The mass
accretion rate is calculated as the mass passing through the sink cell per time step 
of numerical integration (which in physical units is usually equal to 10--20 
days and the total integration time can amount to 2.0~Myr).
The protostar acquires a fraction $\xi$, ranging between 50\%-90\%, of the initial prestellar core's mass, implying that a fraction $1-\xi$  is assumed to have been carried away by outflows\footnote{As mentioned in VB10, the smaller the fraction of core mass ending up as the protostar mass, the larger the disk to star ratio, favoring more disk instability and thus burst accretion.}.

In the present calculations, the structure and evolution of the accreting protostar is consistently
calculated from accreting evolution models (see \S 3). We stress that these calculations take into
account the thermodynamic properties of the gas, i.e., protostellar and background irradiation,
radiative cooling and viscous and shock heating (see VB10). We have conducted simulations, and thus obtained accretion histories, for initial prestellar core masses $\mc$ ranging from 0.06 $\msol$ to 1.85 $\msol$. 
The strongly variable accretion rates obtained in this type of simulations have been shown to reproduce the observed steep $\dot{M}$--$M_{\star}$
relation in class II objects  (Vorobyov \& Basu 2009). 
They also reproduce the observed spread for protostar signatures both in the
$L_{\rm bol}$--$T_{\rm bol}$ and $L_{\rm bol}$--$\mc$ diagrams, including the very low luminosity objects (VELLO's) (Dunham \& Vorobyov 2011). Even more interestingly, they provide an excellent match to these protostellar observed luminosity distributions, 
strongly supporting episodic accretion as the most plausible explanation for
the long-standing luminosity problem.

VB10 and V10 have explored the dependence of the results upon the initial core mass, $\mc$,  and rotational to gravitational energy ratio, $\beta=E_{rot}/E_{grav}$, within a range of values ($\beta = 10^{-3} - 3 \times 10^{-2}$)  typical of observations of dense molecular
clouds (Caselli et al. 2002). They show that high $\mc$ and 
high $\beta$ favor disk fragmentation, because of the increase in the disk mass due essentially to the larger initial core mass or the larger centrifugal radius for the more massive or more rapidly rotating cores. Given that the accretion burst intensity correlates with the disk's propensity to fragment (the bursts are caused by massive fragments merging with the protostar due to gravitational torque interactions with the
spiral arms),  the accretion burst intensity is thus found to
increase with $\mc$ 
(see e.g. Fig. 8 of VB10) and with $\beta$ (see e.g. Figs. 5 and 7 of VB10 and Fig. 6 of V10). 
It is worth stressing that in these simulations, only a combination of a high prestellar core mass, $\mc \simgr 0.7\, \msol$, and a high rotation rate,
$\beta \simgr 5\times 10^{-3}$, can produce burst intensities $> 10^{-4} \msolyr$, as inferred for Fu Ori type objects (HZC11, see \S1). Similar results have been found in 2D and 3D simulations exploring the same issue (Machida et al. 2010, Zhu et al. 2011).

These prestellar core collapse simulations have two major limitations. First, they are limited to 2D, though with an approximate reconstruction of the disk vertical structure. The main reason for such a limitation is that 2D simulations allow a numerical resolution which is presently out of reach with 3D simulations, an important issue to accurately describe the disk structure and evolution, in particular when taking into account the thermodynamics of the gas. The second reason is that 2D simulations enable us to cover many disk orbital periods (with absolute integration times of order 1.0--2.0~Myr), which is not possible so far in 3D, again a major issue when exploring the disk's long-term evolution and stability and the fate of the massive fragments. 
A first set of 3D simulations exploring the stability of the disk during protostar formation has recently been obtained by Machida et al. (2010, 2011). These simulations qualitatively and even semi-quantitatively agree with the 2D ones of Vorobyov \& Basu, exhibiting the same episodic accretion phenomena, with comparable burst intensities. The second limitation of the present simulations is the absence of magnetic field. It is well known that magnetic field plays a crucial role in the prestellar core collapse and can significantly hamper disk fragmentation, at least in ideal MHD calculations (Banerjee \& Pudritz 2006, Price \& Bate 2007, Hennebelle \& Teyssier 2008, Machida et al. 2008, Hennebelle et al. 2011). The ideal MHD simulations of Vorobyov \& Basu (2006) have shown that magnetic fields can significantly reduce the burst intensity, although this effect can be counterbalanced by increasing the initial mass and angular momentum of the cloud prestellar core.
The importance of magnetic field upon the formation of the proto circumstellar disk has received support from confrontation with observations (Maury et al. 2010). Recent resistive MHD simulations, however, suggest that, although considerably smaller than that in an unmagnetized cloud, a circumstellar disk can still form in a strongly magnetized cloud (Inutsuka et al. 2010, Machida et al. 2011), even though the issue does not appear to be completely settled yet (Li et al. 2011). The orientation of the field is also found to have an impact, with departure of the field from alignment with the rotation axis favouring the formation of a disk (Hennebelle \& Ciardi 2009). Therefore, although the exact burst characteristics obtained in the present 2D, hydrodynamics calculations may be questionable from the very quantitative point of view, the general properties of the accretion histories seem to be reasonably sound.

Finally, a limitation of the present calculations is the lack of self-consistent 
treatment of the radiative feedback from the accreting protostar. 
Our numerical hydrodynamics code (VB10) uses accretion and photospheric luminosities 
calculated from the {\it non-accreting} stellar evolution models of D'Antona \& Mazitelli (1997),
which differ from those calculated from {\it accreting} stellar evolution models.
Fully self-consistent simulations require
the coupling of our hydro and stellar evolution codes in one real-time numerical simulation of 
the stellar and disk evolution during the main accretion phase.
Work to produce such self-consistent models is under progress (Vorobyov, Baraffe \& Chabrier, in prep.).

\section{Accreting models for the protostar and brown dwarf evolution}\label {evolution}
\label{setup}

The general energy equation for an accreting object reads (Chabrier et al. 2007a, BCG09):
\begin{eqnarray}
L_{\rm tot} &=&(1-\alpha)\epsilon\frac{GM{\dot M}}{R}+\alpha\epsilon\frac{GM{\dot M}}{R} \nonumber \\
 & - &\int_MT\{(\frac{\partial S}{\partial t})_q-{\dot m}(\frac{\partial S}{\partial m})_t\}\,dm\,+\,L_D \nonumber \\
 \label{eq1}
\end{eqnarray}
The first term of the r.h.s. of eqn.(1) corresponds to the radiated accretion luminosity, $L_{acc}$, the second term to the fraction of accretion kinetic energy absorbed by the protostar; the third term corresponds to the protostar intrinsic entropy rate at constant mass shell $q=m/M$ plus the extra entropy at constant time arising from the accreted matter; the last term denotes the deuterium luminosity arising from both the protostar deuterium content and the one accreted with the infalling material. We  adopt the same assumption as in Hartmann et al. (1997, see their Eqs(3)-(4)) and BCG09, namely that the accretion occurs over a fraction $\delta$ of the stellar surface,
so that 
\begin{equation}
L_{acc} = 4 \pi R^2 \delta F_{\rm acc} , 
\end{equation}
where $F_{\rm acc}$ is the radiative flux averaged over the accreting region.
Models of continuum emission from optical surveys suggest that $\delta \lesssim 10\%$ (Hartmann et al. 1997), so we will assume $\delta=0$ in the rest of the paper, as in BCG09, for sake of simplicity.  Further assuming, as in Hartmann et al. (1997), that there is a negligible energy transfer between the non accreting stellar surface and the region where accretion occurs, the stellar luminosity reads

\begin{equation}
L_\star =  4 \pi R^2(1-\delta) \sigma \te^4 \approx 4 \pi R^2 \sigma \te^4
\end{equation}
The impact of reducing the star radiating surface can be inferred from  the  study of Chabrier et al. (2007b), who explored  the effect of magnetic spot coverage.  These authors found that a surface spot coverage of 20\% affects the radius evolution of young brown dwarfs by less than  5\% (see Fig. 3 of Chabrier et al. 2007b). Consequently, for $\delta \lesssim 10\%$ or so,  we do not expect a significant effect on the evolution/radius compared with   the much larger impact due to accretion, as shown below.

In Eq(\ref{eq1}), the factor $\epsilon$ is the amount of internal energy per unit mass brought up by the accreting material ($\epsilon \lesssim 1$ for spherical accretion while $\epsilon  \lesssim 1/2$ if accretion occurs from a thin disk at the object's equator (Hartmann et al. 1997). As in BCG09, a value $\epsilon=1/2$ will be assumed in the present calculations. Finally, the factor $\alpha\le 1$ is the fraction of accretion energy absorbed by the protostar at the accretion shock, while $(1-\alpha)$ is the fraction radiated away. 
The evolutionary models are based on the same input physics and general framework as described in BCG09 and BC10, except for three main differences, as described in the next subsections.  

\subsection{Accretion rates}
As mentioned above, the accretion rates are not arbitrary, but 
derived  from the hydrodynamics simulations of prestellar core collapse mentioned in 
\S \ref{accretion}. 
A supercritical shock at the protostar surface corresponds to $\alpha=0$, implying that in such a case all the accretion energy is radiated away and does not contribute to the protostar internal heat content. Such supercritical shock conditions have been found in 1D complete RHD calculations of the first collapse (first Larson's core) of a 1 $\msol$ and a $0.01 \, \msol$ prestellar core, both when using a grey approximation (Commer\c con et al. 2011) and a more accurate multigroup frequency description for the opacity (Vaytet et al. 2012). There is no garantee, however, that the same results still hold for the second collapse. Work to explore this issue is under progress.
As shown in Hartmann et al. (1997) and BCG09, the value $\alpha_{crit}\approx 0.1$-0.2 corresponds to a critical value.  While
a value $\alpha<\alpha_{crit}$ leads to a {\it contraction} of the accreting protostar compared to the non-accreting counterpart at same mass and age, a value $\alpha>\alpha_{crit}$ means that enough energy is absorbed by the accreting protostar to balance or even dominate its contraction, leading to a null effect or an {\it expansion}  compared to the non-accreting counterpart. The first case refers to what is denominated "cold" accretion, meaning that the entropy of the infalling material is
significantly  lower then the one of the protostar, while the second case corresponds to "hot" accretion, as briefly mentioned in \S\ref{intro}.

\subsection{Initial protostar mass} \label{ini}
In BCG09, the 
{\it protostar initial masses}, that will be denoted $\mi$, were chosen to range from 0.001 to 0.1 $\msol$. This range was meant to bracket the typical value  of the first Larson's core, as found in 1D (Larson 1969, Masunaga et al. 1998) and 3D  (Commer\c con et al. 2010, Tomida et al. 2010) numerical simulations of the first collapse. A more relevant value for the initial mass of the protostar, however, is the one typical of the {\it second} adiabatic Larson's core, which forms when the  central density reaches values around $\sim 10^{-2}$ g cm$^{-3}$, characterising the  so-called stellar density in cloud collapse calculations (Larson 1969, Masunaga \& Inutsuka 2000). Note, however, that this mass remains ill-determined, as no 3D simulations of the second collapse including the appropriate detailed physics (EOS, radiation hydrodynamics, magnetic field) have been carried out yet. We thus must rely on existing calculations or order of magnitude estimates.


A crude estimate for the second core's initial mass can be inferred as follows. Assuming a polytropic EOS, $P=K_n\rho^{1+1/n}$, the structure of the second core can be estimated as a scaled down version of the first core (Masunaga et al. 1998), i.e. from a second-order Lane-Emden equation. This yields for the central density and pressure:

\begin{equation}
\rho_c=A_n\,{\bar \rho}=\frac{3A_nM}{4\pi R^3}\,\,;{\hskip .5cm} P_c=W_n\,\frac{GM^2}{R^4},
\end{equation}
where $M$ and $R$ denote the mass and the radius of the second core, respectively, and $A_n$ and $W_n$ are the appropriate polytropic constants (Chandrasekhar 1957). Assuming a perfect gas EOS yields for the central temperature
\begin{equation}
T_c=\frac{\mu m_H}{k_{\rm B}}\frac{P_c}{\rho_c}=\frac{4\pi G\mu m_H}{3k_{\rm B}}\frac{W_n}{A_n}\frac{M}{R},
\end{equation}
where $\mu$ denotes the mean molecular weight, $m_H$ the atomic mass unit and $k_{\rm B}$ the Boltzmann constant.
The second hydrostatic core will form when all molecular hydrogen is dissociated, i.e. when the temperature reaches about 5000 K (Saumon et al. 1995). This yields for the second core minimum mass:
\begin{eqnarray}
M_i&\equiv& M_{min}=\frac{3k_{\rm B}}{4\pi G\mu m_H}\frac{A_n}{W_n}{T_{min}}{R}\nonumber\\
\Rightarrow 
({M_{i}\over M_\odot})&\simeq& 4.0\times 10^{-4} \,(\frac{\mu}{1.0})^{-1}(\frac{T_{min}}{5000\,{\rm K}})(\frac{R}{R_\odot})\nonumber\\
&\simeq& 8.65\times 10^{-2} \,(\frac{\mu}{1.0})^{-1}(\frac{T_{min}}{5000\,{\rm K}})(\frac{R}{{\rm A.U.}}),\nonumber\\
\label{core}
\end{eqnarray}
where the values have been estimated for a fully ionized monoatomic polytropic gas $n=3/2$. Taking $R\approx 1\,\rsun$ for the second core's radius, as inferred from the location of the first accretion shock in numerical simulations of the second collapse (Larson 1969, Masunaga \& Inutsuka 2000), eqn.(\ref{core}) yields about half a Jupiter mass for $M_i$. 
The most recent numerical determination of the minimum mass for opacity-limited fragmentation by Boyd \& Whitworth (2005) yields 3 $\mjup$. Whitworth \& Stamatellos (2006) have derived a minimum mass for star formation, denoted as minimum mass for primary fragmentation, in the range 1-4  $\mjup$. 3D hydrodynamics or MHD barotropic numerical simulations of the collapse of molecular clouds to stellar densities (Bate 1998, Intsuka et al. 2010, Machida et al. 2010, 2011) yield a second core mass $\mi \simeq 1\,\mjup$.

It is unclear, however, to which extent variations in the initial cloud prestellar core conditions will affect this value.
The thermodynamics of the collapsing gas, for instance, may have some impact on the final (protostar) core mass. For a polytrope, the internal entropy is related to the central pressure and density as $S_{int}\propto \ln (\frac{P_c}{\rho_c^{5/3}}) \propto \ln ({R}{M}^{1/3})\propto \ln {M}^{4/3}$, where we have made use of eqn.(3) (e.g. Stahler 1988). Assuming a perfectly adiabatic collapse, this implies a dependence of the second core mass upon the entropy  $M_i\propto \exp\{\frac{3}{4}(S_{int}-S_0)\}$, where $S_0$ denotes the entropy at some initial stage before the onset of the second collapse. Because of the thermal properties of the gas, however, the collapse is not expected to be perfectly adiabatic, as part of the contraction work ($PdV$) will be evacuated by radiative cooling, implying $dQ\ne 0$. 
The gas may also get heated by the accretion luminosity onto the central object. Indeed, variations of the entropy at the first core stage are found in RHD collapse simulations (Commer\c con et al. 2011). These variations, however, remain modest, $\lesssim 5\%$. Although the impact at the second core stage remains to be explored, this suggests that variations of the second core mass due to the collapsing gas thermal properties are likely to remain at the few percents level, except in particular situations where the gas has been heated significantly due e.g. to strong radiative feedback from a nearby massive star. Similarly, variations in the initial rotation of the cloud prestellar core are also found to barely affect the {\it initial} mass of the protostar (Machida, private communication), suggesting a limited impact of rotation on the second core mass. Magnetic braking, however, is found to increase the limit of fragmentation and thus the minimum core mass during the first collapse (Commer\c con et al., 2010). Even though, again, no calculation exists for the second collapse yet, the impact of magnetic field at this stage is likely to remain weak, as most of the flux is expected to get lost by ambipolar diffusion during the first collapse.
In summary, it seems difficult to significantly modify the initial mass of the protostar, $M_i$, 
even though this issue remains to be examined by dedicated calculations of the second collapse. 

Given these results, and awaiting for more accurate determinations of the second core mass from 3D full RMHD simulations of the second collapse, we have elected in the present calculations  to take for the value of the initial protostar mass, at the very beginning of the accretion process,  a value $\mi=1\,\mjup$, with possible variations up to $5\,\mjup$. It is essential at this stage to stress  that  HOK11 consider {\it only one value} for this seed mass\footnote{Note that the definition of the initial mass used in our calculations and in the calculations of HOK11 is the same, representing the initial mass for the stellar evolution calculations.}, namely $\mi=10 \mjup$.
As mentioned above, such a value is representative of the typical  {\it first} Larson's core, but is likely to be quite a too extreme value to examine the early evolution of low-mass stars and brown dwarfs under standard cloud conditions.

\subsection{From cold to hot accretion} \label{hot}

The third difference w.r.t BCG09 is that only two values for $\alpha \ne 0$ were considered in these calculations, namely $\alpha$= 0.2 and $\alpha$= 1 (see Table 1 of BCG09).
The value $\alpha$= 0.2 was found to be the one for which the protostar's accretion induced contraction was balanced nearly exactly by internal heating induced expansion, defining the limit between "cold" and "hot" accretion. In the present study, we explore more values for $\alpha$, between 0 and $\sim$ 0.2, in order to quantify the impact of even a modest fraction of absorbed accreting energy upon the protostar evolution. Note that such finite values for $\alpha$ ($\alpha \ne 0$) are considered only for sufficiently high accretion rates, namely rates exceeding some critical value $\dot{M}_{\rm cr}\ge10^{-5} \msolyr$, as inferred from Fu Ori events (see \S 1). 

 \section{The HRD luminosity spread} \label{HRD}
 
A significant spread in the HRD for  low-mass objects has been observed in many young, $\sim$Myr old, clusters and star forming regions. The exact amount of the spread is difficult to assess, as it can vary between the regions and is affected by various sources of uncertainties. The typical $L$-$\te$ range, however, corresponds to the region bracketed by the 1 Myr and $\sim 10$ Myr  Baraffe et al. (1998) isochrones for non-accreting objects in the LMS and BD domain, although some observations suggest an even larger spread (see e.g. Bayo et al. 2011 for the Lambda Orionis star forming region). This typical observational spread is illustrated in Fig. 1 of BCG09. In the present section, we will show that episodic accretion can lead to such a typical spread after 1 Myr.

 \subsection{Lower part of the HRD} \label{small}
 Using the conditions described in \S \ref{evolution}, we first explore the impact of accretion for objects in the cool part of the HRD, in the region $\te \simle 3000$ K. HOK11 claim that, in this domain,  it is not possible to produce a luminosity spread, except if one adopts unrealistically small values for the initial
  radius (thus entropy) of the protostar (see Figs. 6 and 7 of HOK11).  
  
 \subsubsection{Variations of the initial protostellar mass} \label{ini}
 
\begin{figure}[h!]
\begin{center}
\includegraphics[height=10cm]{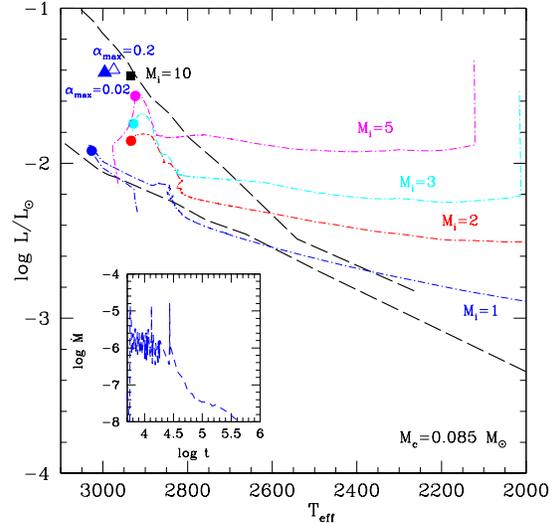}
\caption{Evolution of accreting sequences for initial seed masses $\mi$ varying from 1 to 10 $\mjup$ and accretion rates 
derived from the collapse of a prestellar core of mass $\mc = 0.085 \msol$ and $\beta=2.3\times 10^{-2}$. Dashed-dotted lines  correspond to $\alpha$=0. Filled circles on each coloured curves give the position at 1 Myr. All tracks terminate at a few Myr. The black filled square indicates the position at 1 Myr of a sequence starting with $\mi$=10 $\mjup$ and $\alpha$=0. The triangles indicate the position at 1 Myr of models calculated with $\mi$=1 $\mjup$, $\amax$=0.02 (filled triangle) and $\amax$=0.2 (open triangle) (evolutionary tracks are not shown for these sequences for sake of clarity). The two long dashed (black) curves are the 1 Myr and 10 Myr isochrones of Baraffe et al. (1998) for non accreting models. The inset shows the accretion rate (in $\msolyr$) as a function of time (in yr). 
}
\label{fig1}
\end{center}
\end{figure}

 Figure \ref{fig1} shows 
 sequences of evolution obtained from the collapse of an 
initial prestellar core $\mc$ =0.085 $\msun$, with $\beta=2.3\times 10^{-2}$, for various values of $\mi$ between 1 and 5 $\mjup$, for $\alpha=0$. The inset in the figure illustrates the accretion history of the collapsing parent cloud core. Note that, for such a small initial core mass, the accretion rate barely exceeds $10^{-6}\,\msolyr$, as mentioned in \S \ref{setup}, with only two weak accretion
bursts which occur in the early phase due to a high (yet within the observational upper limit) value of $\beta$.
We stress again that the accretion history (bursts, rates) during the evolution of the protostar 
is the one obtained from the core collapse calculations, insuring complete consistency of the calculations. All evolutionary sequences terminate at $\sim$ 1-2 Myr and the positions at 1 Myr are indicated by the solid circles. At this age, the final object produced along these sequences is a brown dwarf with a final mass $\mf=$0.065-0.07$\msol$, depending on $\mi$. As seen in the figure, a range of initial protostar masses between 1 and 5 $\mjup$ does lead to the sought typical luminosity spread after 1 Myr, for $\alpha=0$. As discussed in \S\ref{setup}, such a variation among the protostar initial masses at the second core stage can not be excluded at the present time. As seen in the figure, protostar initial masses as small as $\mi\simeq 1\,\mjup$ are required in order to produce objects as faint as the ones corresponding to the 10 Myr non-accreting isochrone\footnote{Such faint objects are indeed observed in few Myr old clusters (see BCG09 and references to observations therein).}. 
Since, assuming uniform core-to-star mass conversion due to outflows, as suggested by observations (e.g. Andr\'e et al. 2010), small protostar masses can only be produced by small initial cloud prestellar core masses, this in turn implies the existence of small ($\lesssim 0.1 \,\msol$) prestellar core masses. Such cores are starting to be discovered with Herschel (Andr\'e et al. 2010, K\"onyves et al. 2010) and are predicted by analytical gravo-turbulent theories of cloud fragmentation (Padoan \& Nordlund 2002, Hennebelle \& Chabrier, 2008, 2009).  As seen in the figure, the obvious reason why HOK11 could not succeed producing a spread in the HRD for objects 
with final temperature $\te \simle 3000$ K, i.e. $M_f \simle 0.1 \msol$, is that their initial protostar mass is set up to be $\mi=$10 $\mjup$, which indeed leads to a final object close to the 1 Myr non-accreting isochrone. In order to verify this statement, we have performed a calculation for $\mi=$10 $\mjup$ (for the same prestellar core mass $\mc$=
0.085$\msun$): we indeed obtain an object with $\te \sim 3000$ K and a position at 1 Myr very close to the 1 Myr non-accreting isochrone,  similar to the results of HOK11 (see black filled square in Fig. \ref{fig1}). 
This  confirms the fact that we can reproduce the calculations of HOK11, if we use the same initial conditions, and that the very reason for their disagreement with the BCG09 results is the use of one single value,  $\mi=$10 $\mjup$, for the initial protostar seed mass, a very likely too large value for the typical {\it second} Larson's core mass, representative of this seed mass, as mentioned earlier.
An other interesting effect illustrated in the figure is that, for the present prestellar core mass $\mc$ =0.085 $\msun$, the impact of early accretion is significant even after 1 Myr for initial masses $\mi \lesssim 5 \mjup$, while the effect vanishes for larger initial masses after about this time. A quantitative explanation for such a behaviour, which involves deuterium burning, is given in Appendix A.
A comparison between the final mass of accreting vs non-accreting objects for a given ($L$, $\te$) location in the HRD  will be given in \S \ref{IMF}.

Figure \ref{fig1}  illustrates the effect of the variation of  the initial protostar mass $\mi$ on the evolutionary sequence, for a given prestellar core mass. As shown in BCG09 and as just mentioned, for a given accreted mass $\Delta M$, the smaller $\mi$, thus the smaller the initial binding energy, the larger the impact of accretion on its structure (see explanation in Appendix A). These calculations show that, considering different values of $\mi$  between 1 and  5 $\mjup$ naturally produces the spread in the cool part of the HRD. 

We stress again that, as stated in BCG09,  the impact of accretion on the structure of an object depends essentially on the {\it total amount of accreted mass}, $\Delta M$, since the initial protostar very formation, and depends only weakly on the detailed accretion history, i.e constant or monotonic vs episodic accretion (see \S 3.2 of BCG09). This has been confirmed by HOK11.  
Besides the reasons mentioned in \S1, the obvious reason in favor of {\it episodic} accretion is that it takes a significantly shorter time for a star to build up its mass from an initial protostar mass through a few intense accretion bursts than from a steady, or more likely rapidly decreasing (see the inset in the figure) "standard" accretion rate. 

Fig. \ref{fig1} also illustrates the impact of a finite value of $\alpha$, i.e. of some fraction of accreting energy absorbed by the protostar, on its evolution. As mentioned in \S\ref{hot}, we define $\dot{M}_{\rm cr}$ as some critical accretion rate above which  $\alpha$ is switched from 0 (for $\mdot < \dot{M}_{\rm cr}$) to some finite value $\alpha_{\rm max}$ (for $\mdot > \dot{M}_{\rm cr}$). Based on the Fu Ori event mentioned in \S1, we arbitrarily define this critical rate as $\dot{M}_{\rm cr} =10^{-5}\msolyr$.  As seen in the figure, for a protostar initial mass $\mi=1\,\mjup$, 
$\alpha_{\rm max}$  $\sim$ 2 \%, i.e. a tiny fraction of accreting energy absorbed by the protostar, is sufficient to counterbalance the contracting effect of mass accretion, yielding eventually a position in the HRD  for the final object after 1 Myr similar to that of its non-accreting counterpart at same mass and age. Interestingly enough, a significantly larger value, $\alpha_{\rm max}=0.2$, leads basically to the same result for such small initial masses. 
 As explained in  \S 3.1 of BCG09, this stems from the fact that, for $\alpha \ne 0$ (even for values as small as 0.02), the extra energy absorbed during the few bursts with $\mdot > \dot{M}_{\rm cr}$ yields an evolution which proceeds at higher luminosity and larger radius than the non-accreting counterpart, which implies shorter thermal timescales $\tkh = GM^2/(RL)$. After the last intense burst (at $\log \, t=4.5$ in Fig. \ref{fig1}), for instance, the thermal timescale of the accreting objects with $\alpha \ne 0$ is shorter than the accretion timescale, i.e. $\tkh < \tmdot=M/\dot M$. This is  in stark contrast with the case $\alpha=0$, for which $\tkh > \tmdot$ for the same burst intensities. As explained in Chabrier et al. (2007a) and BCG09, in the first case, the object quickly contracts and converges towards a location in the HRD very close to that of the non accreting counterpart. On the opposite, in the second case the thermal timescale is much longer, of the order of a few Myr, and only at this stage will the initial conditions due to early accretion be forgotten.
Such high accretion episodes leading to $\alpha\ne 0$, however, should be rare in small-mass collapsing $\mc$ cores, at least according to our choice of $\dot{M}_{\rm cr}$. Indeed, $\dot{M}$ hardly exceeds the value
$10^{-5}\msolyr$ for such cores, even for large values of $\beta$, as seen in the inset of Fig.~\ref{fig1} (see also VB10). Accretion  bursts in low-mass models occur
only for cores with rather high values of rotational support, $\beta\ga2\times 10^{-2}$, 
because high disk-to-star mass ratios are needed to trigger a burst.
Note, however, that even a small number of such events is sufficient to significantly alter the radius, thus the luminosity of the accreting object. Had we adopted a slightly larger value of $\dot{M}_{\rm cr}$ (say $3\times 10^{-5}~\msolyr$), our criterion for a non-zero value of $\alpha$ would never have been fulfilled for the $\mc$ model shown in Fig. \ref{fig1}. In that case, only a variation among the initial protostar masses, $\mi$, could lead to the typical luminosity spread. Note that, according to the BCG09 and present calculations, large values of $\alpha \simgr 0.2$ would imply the existence of very bright very-low-mass Fu Ori like events, located above and on the rightmost side of  the 1 Myr non-accreting isochrone. The lack of such observations, so far,  in the very low-mass domain of the HRD thus suggests that either they last a very short time, or they are intrinsically very rare, or even absent, indeed suggesting very small values of $\alpha$ in this (low-mass) domain. This diagnostic, if confirmed in the future, would thus confirm on observational grounds the results obtained in the present and VB10 collapsing core calculations, namely that (i) the burst frequencies and intensities directly correlate with the initial core mass, and (ii) that hot accretion (defined as $\alpha \simgr 0.2$) is unlikely to occur for low-mass cores, due to the lack, or at least the rarity of intense accretion bursts.

 \subsubsection{Variations of the initial prestellar core mass} \label{mc}
 
\begin{figure}[h!]
\begin{center}
\includegraphics[height=10cm]{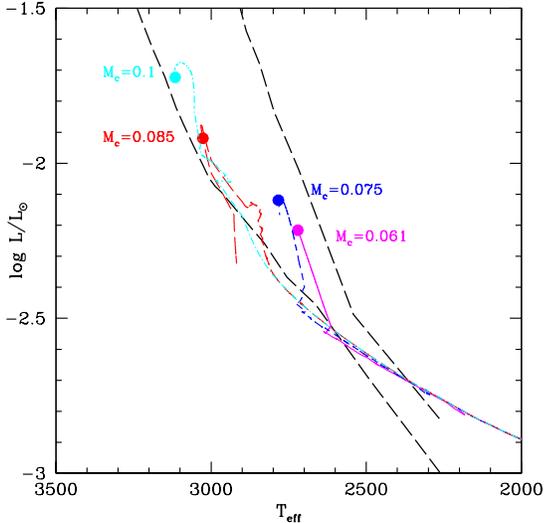}
\caption{Evolution of accreting sequences for four different initial prestellar core masses, namely $\mc=$ 0.061 (solid, magenta), $\mc=$ 0.075 (dash, blue), 0.085 (long-dash, red), and 0.1 (dash-dot, cyan) $\msol$, starting from  $\mi$=1 $\mjup$, with consistent respective accretion histories, for $\alpha$=0. Filled circles give the position at 1 Myr. The two long dashed (black) curves are the same as in Fig.\ref{fig1}. 
}
\label{fig1bis}
\end{center}
\end{figure}

In this subsection, we explore a different strategy. Given that the second Larson's core mass is essentially fixed by the entropy of the gas at the end of H$_2$ dissociation, i.e. by the H$_2$ temperature dissociation and the density at which the gas becomes opaque (the aforementioned typical stellar density), and that this core mass should not depend much on the initial cloud conditions, the protostar initial mass will always have basically the same value. Fig. \ref{fig1bis} illustrates in that case the evolutionary tracks obtained for the same initial protostar mass, $\mi$=1 $\mjup$, for accretion histories obtained from four different  initial prestellar core masses, namely $\mc=$ 0.061, 0.075, 0.085 and 0.1 $\msol$, for $\alpha$=0. The four sequences produce objects with final mass 0.03, 0.04, 0.065 and 0.09 $\msol$ respectively. As seen in the figure, the smallest $\mc$ produce  objects at 1 Myr with a position in the HRD intermediate between the 1 and 10 Myr isochrones, while the two largest $\mc$ yield objects with luminosities corresponding to the 10 Myr non-accreting isochrone. As explained in Appendix A, these two quantitatively and even qualitatively different behaviors reflect the balance between the protostar's binding energy and deuterium burning nuclear energy, for objects massive enough to ignite D-fusion.  These results show that, for a fixed protostar initial mass $\mi$=1 $\mjup$, and within the effective temperature range  $\te\approx 2500 - 3000$ K, a variation of the cloud initial prestellar core mass can produce a luminosity and $\tef$ spread, populating again the region in the HRD between the 1 and 10 Myr non-accreting isocrones.
Needless to say, a combination of these results with the ones obtained in the previous subsection for a given prestellar core mass and a variation of the values of $\alpha$ will produce a nearly homogeneous distribution of young, $\sim 1$ Myr old stars and BDs over the entire region between these two non-accreting isochrones. The case of BD's below  0.04 $\msol$ will be examined in \S 7.

To summarize the present section, we can say that a luminosity spread in the lower, coolest part of the HRD ($\te \simle 3500$ K, $M \simle 0.1 \msun$) similar to the observed one can be produced 

(i) either by starting with a spread of initial protostellar masses in the 1-5 $\mjup$  range,  for $\alpha$=0, 

(ii) or by starting with an unique initial protostellar mass, $\mi \simeq 1\,\mjup$, but considering collapsing low-mass cloud prestellar cores of different masses -  as indeed expected from 
the core mass distribution resulting from the parent cloud collapse - with some variation of $\alpha$ between 0 and a few percents.

Or, obviously, by a combination of both scenarios.
 Note, however that, in both cases, it is necessary to start with protostar initial masses {\it as small as $\sim 1\,\mjup$} to obtain LMS or BDs as faint at 1 Myr as non-accreting 10 Myr old objects of same mass. The unambiguous observation of such faint low-mass objects in young ($\sim 1$ Myr) clusters would thus indicate that the minimum mass for star formation indeed extends down to about this limit. In contrast, the demonstration that star formation can never produce such small initial protostar masses would suggest that either some physics is missing in the present calculations, or indeed the luminosity spread in the low mass part of the HRD is not, or at least not entirely due to early accretion in the protostar phase, or that the observations are questionable.

 \subsection{Upper part of the HRD} \label{upper}
 
\begin{figure}[h!]
\begin{center}
\includegraphics[height=10cm]{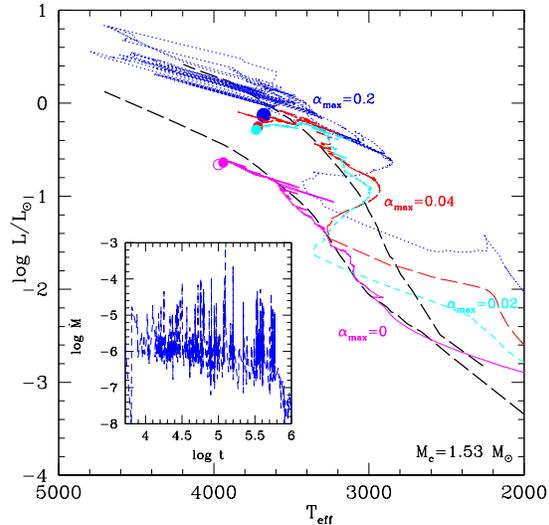}
\caption{Evolution of accreting sequences with $\mi$ =1 $\mjup$ and different values of $\alpha$, with accretion rates (plotted in the inset in $\msolyr$ versus time in yr) derived from the collapse of a prestellar core of mass $\mc = 1.53 \msol$ and $\beta=1.3\times 10^{-2}$. 
 Filled circles on each coloured curves give the position at 1 Myr.  The (magenta) open circle close to the filled circle corresponds to the position at 1 Myr of a sequence starting with $\mi$ = 5 $\mjup$ and with $\alpha_{\rm max}$=0. The two long dashed (black) curves are the same as in Fig. \ref{fig1}.}
\label{fig2}
\end{center}
\end{figure}

 We now focus on the upper, hotter part of the HRD, namely final objects with $\te > 3000$ K, up to $\sim$ 1 $\msol$. Always assuming a roughly mass-independent fraction of  core accreting envelope removed by outflows, such large masses must be issued from larger  cloud initial prestellar core masses. Figure \ref{fig2} displays sequences of evolution obtained for an initial prestellar core $\mc$ =1.53 $\msun$, with
$\beta=1.3\times10^{-2}$, starting from a protostar initial mass $\mi = 1\, \mjup$, for different values of $\alpha_{\rm max}$ (we recall that $\alpha$ is switched from 0 to $\alpha_{\rm max}$ {\it only} during burst events exceeding $\mdot \ge \dot{M}_{\rm cr}=10^{-5}\msolyr$). 
The prestellar core collapse calculations were performed up to $t$=1.5 Myr. At this age, the star has a mass $M$=0.75 $\msun$ and still accretes at a rate $\mdot \sim 8 \times 10^{-8} \msolyr$. At 1 Myr, the central object has a mass $M$=0.71 $\msun$. 
The results show that variations of $\alpha_{\rm max}$ in the range 0-0.2 produce the typical observed luminosity spread in the HRD. Note in passing the incredibly erratic evolutionary path for $\alpha_{\rm max}=0.2$. These tracks, as well as the ones illustrated in the previous figures, highlight the complex early evolution of accreting protostar until they reach their final mass, quite different from the usual constant-$\te$ Hayashi track typical of low-mass non accreting objects. For illustration, we also display the position at 1 Myr  of a sequence starting with an initial mass $\mi$ = 5 $\mjup$ and $\alpha_{\rm max}$=0 (empty circle).
As seen, starting with $\mi = 1$ or $5 \,\mjup$ leads basically to the same results, a consequence of the large amount of total mass accreted and the resulting large binding energy of the protostar.  Indeed, for such sequences, the energy release from deuterium burning can never overcome the gravitational energy increase (as explained in Appendix A), and the effect of accretion is the same, independently of the initial mass.                  
A similar spread  was obtained, with cold accretion, in BCG09, for larger values of $\mi$ 
(see discussion in \S \ref{ini}). As illustrated in Figure \ref{fig2}, adopting values for $\mi$ more consistent with the ones expected for the second Larson's core, the only possibility to counterbalance the contracting effect of mass accretion on the protostar and to produce a luminosity spread is to assume a spread in the values of $\alpha$ between 0 and $\sim 0.2$, i.e. some variations in the shock conditions on the protostar. 
This is consistent with the accretion histories obtained in the present and VB10 calculations, as well as in 3D simulations (Machida et al. 2010). Indeed, in contrast to the case with small prestellar 
cores,  the accretion rate for the {\it larger} cores (even with moderate values of $\beta\ga 5\times 10^{-3}$) often exceeds our critical value 
$\dot{M}_{\rm cr}=10^{-5} \msolyr$ and  even $10^{-4}~\msolyr$ during the most intense bursts. It is intuitively expected that larger accretion rates, thus larger ram pressures and hotter accreting material, are more likely to affect the internal heat content of the initial protostar, fixed by the second Larson's core adiabat. In Appendix B, we provide some analytical estimates of the properties of the accretion shock as a function of the accretion rate which provide some physical foundation for such larger values of $\alpha$ for higher mass objects.
 
 \section{Fu Ori bursts}
\begin{figure}[h!]
\begin{center}
\includegraphics[height=10cm]{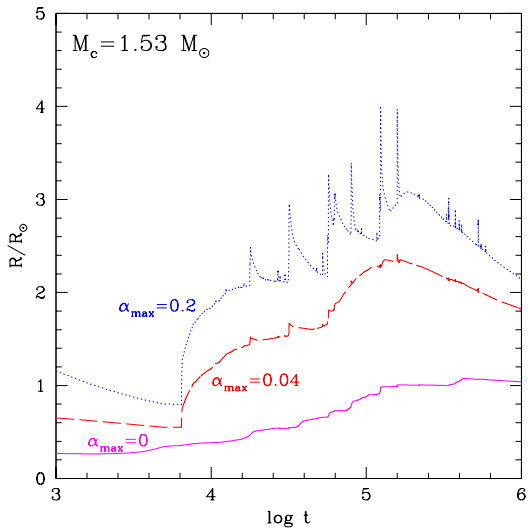}
\caption{Evolution of the radius (in solar radius) with time (in yr) for accreting sequences with $\mi =1 \mjup$ and different values of $\alpha_{\rm max}$ (same sequences as Fig. \ref{fig2}).}
\label{fig3}
\end{center}
\end{figure}

 As displayed in the inset of Fig. \ref{fig2}, the calculated accretion rates corresponding to the collapse of a prestellar core of mass $\mc$=1.53 $\msun$ exhibit strong variations with time, with violent bursts reaching values characteristic of Fu Ori-like bursts. As mentioned above, such high rates are never reached during the collapse of the smaller cores investigated in \S \ref{small}. In case of strong accretion, approaching or exceeding the hot accretion limit, the most intense bursts strongly and durably affect the structure of the protostar, provoking its rapid expansion followed by a rapid contraction as soon as the burst stops. This is illustrated in Fig. \ref{fig3}, which portrays the evolution of the radius for protostars of same initial mass $\mi =1 \mjup$ and final mass $\mf=0.75 \msun$, accreting the same amount of mass $\Delta M$, but for different values of $\alpha_{max}$, namely 0, 0.04 and 0.2. The most intense accretion burst for this sequence reaches $ 7.5\times 10^{-4} \msolyr$ at an age of 0.12 Myr, when the protostar has built up a mass of $\sim 0.35\, \msol$. Despite the extremely short duration of the burst ($\sim$ 20 yr), typical of Fu Ori bursts, absorption of 20\% of the accreting energy by the protostar yields an expansion of up to a factor 2 in radius (blue dotted curve in Fig. \ref{fig3}), reaching $\sim$ 4 R$_\odot$. Most interestingly, this sequence matches remarkably well the properties of Fu Ori derived by HZC11, namely a central star mass $\sim$ 0.3 $\msol$, an inner disk radius $\sim$ 5 R$_\odot$ and an accretion rate $\mdot \sim 2\times 10^{-4} \msolyr$. Our calculation thus explains the factor $\sim$ 2 increase in the stellar radius inferred from the spectral energy distribution  of Fu Ori objects. As mentioned in \S \ref{accretion}, such high burst intensities require massive disks, very prone to strong and 
frequent episodes of gravitational instability and fragmentation, and thus massive (large  $\mc$) 
prestellar cloud cores. Such cores do not need to rotate extremely fast and moderate values of $\beta\ga 5\times 10^{-3}$ are sufficient to
produce multiple bursts. However, further decreasing $\beta$ would result in disks of too small size
(due to the corresponding decrease in the centrifugal radius)
that are unable to fragment  and such models would
stop  producing bursts.  This is indeed what is found in the hydrodynamics collapse calculations (see
e.g. VB10 and V10). 
 
\section{The global scenario} \label{discussion}

The results presented in the previous sections suggest the following global scenario for the early evolution of accreting protostars and proto brown dwarfs. Objects in the LMS/BD mass regime form from the collapse of prestellar cores of various masses $\mc$, as produced by the so-called core mass function (CMF), and various initial rotational to gravitational energy ratios  $\beta$ (and magnetic flux intensity as well). There is indeed ample evidence, in particular with the recent results of the Herschel mission, that the bulk of the stellar initial mass function (IMF) resembles the CMF and thus that most stars/BDs should form from the collapse of their parent core (Andr\'e et al. 2010, K\"onyves et al. 2010), in agreement with predictions of analytical theories of gravoturbulent fragmentation (Padoan \& Nordlund 2002, Hennebelle \& Chabrier 2008, 2009, Chabrier \& Hennebelle 2010, 2011).
Initial protostar seed masses, resulting from the second collapse, are assumed to be of the order of $\mi \sim$ 1 $\mjup$, as expected from analytical estimates and as found in existing numerical simulations. Some variations in $\mi$, however, depending on the initial cloud conditions (e.g. magnetic field intensity or orientation) are presently not excluded.  Further studies are definitely needed to explore this issue. 
The larger and the more rapidly rotating the initial cloud core, the more massive the centrifugal disk mass around the central protostar, and the more intense and frequent the accretion burst episodes. As shown above, the combination of high $\mc$ and/or large $\beta$ can produce Fu Ori type bursts ($\mdot > 10^{-4} \msolyr$) and explain Fu-Ori properties (accretion rate, mass, radius). A prediction of such a scenario is that there should be a dearth of Fu-Ori like systems for very low-mass objects (i.e mostly for masses in the BD regime). Assuming, as intuitively expected and as supported by the analytical arguments presented in Appendix B, that there is a correlation between $\alpha$, the fraction of accreting energy absorbed by the protostar,
and the burst intensities, we predict the accretion process onto the central object to be "cold-like" ($\alpha$=0 or at most a few percents) for low accretion rates ($\mdot \lesssim 10^{-5} \msolyr$ or so), {\it i.e} for prestellar cores with small  $\mc$ {\it or} small $\beta$, and to become "milder"  ($0\le \alpha \lesssim0.2$) with increasing $\mc$ and $\beta$, possibly experiencing eventually some episodes of intense "hot" accretion bursts ($\alpha \gtrsim 0.2$), characteristic of Fu-Ori like events. Such a link between $\alpha$ and accretion rates remains to be demonstrated on robust grounds, and will be briefly addressed in \S \ref{conclusion}.
 
 As mentioned above, for low mass prestellar cores, for which burst events do not occur and the accretion rate remains weak ($\ll 10^{-5} \msolyr$) most of the time, the accretion process is predicted to always be essentially "cold-like". The spread of luminosities after $\sim$ 1 Myr in the cool part of the HRD, resulting in an apparent age spread,  can be explained either by variations of the protostar initial mass within the range $1\,\mjup\lesssim\mi \lesssim 5\,\mjup$, or, for a fixed initial mass $\mi \sim 1 \mjup$, by a distribution of initial prestellar core masses and some modest variations of $\alpha$ ($0\le \alpha\lesssim 2\%$). 
For more massive prestellar cores, the sought luminosity spread is best explained, for fixed protostar initial mass, by a larger spread in $\alpha$ values, up to 20\% or so, as mentioned above.  It is worth mentioning, at this stage, that there is not necessarily a direct correlation between the intensity of the bursts and the degree of luminosity spread. As explained in Sect.4.1.1 and appendix A (see also BCG09), the impact of accretion depends essentially on the fact that the accretion timescale, $\tmdot=M/{\dot M}$ , is shorter or not than the thermal timescale, $t_{KH}$. As shown in the previous section, very intense bursts may lead to situations such that $t_{KH}< \tmdot $ with large values of $\alpha$ (not mentioning the role of D-burning), yielding very small or no luminosity spread, whereas more moderate accretion rates could lead to the opposite situation. It also depends on the initial core mass, $\mc$ (see  appendix A).

\begin{figure}[h!]
\begin{center}
\includegraphics[height=10cm]{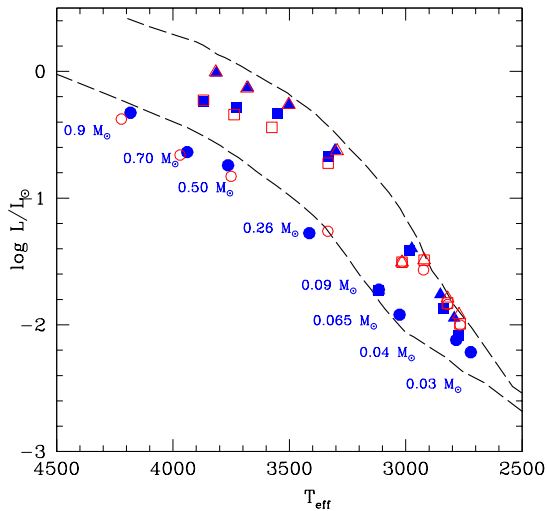}
\caption{Illustration of the luminosity spread in the HRD obtained for various sequences of evolution calculated for prestellar core masses  $\mc$ = 0.061, 0.075, 0.085, 0.1,
 0.38, 0.92, 1.53 and 1.85 $\msun$, leading to objects in the mass range 0.03 - 0.9 $\msol$ at 1 Myr, as indicated in blue on the left hand side of the symbols. The symbols give the position at 1 Myr. Solid (blue) symbols correspond to $\mi$=1 $\mjup$; open (red) symbols correspond to  $\mi$=5 $\mjup$. Circles: $\alpha_{\rm max}$=0; squares: $\alpha_{\rm max}$=0.02; triangles: 
$\alpha_{\rm max}$=0.2. Black dashed curves are the same as in Fig. \ref{fig1}.}
\label{fig4}
\end{center}
\end{figure}

This global scenario   leads to a significant luminosity spread over the entire low mass part ($\le 1\,\msol$) of the HRD {\it at 1 Myr}, as illustrated for a few cases in Fig. \ref{fig4} for various initial prestellar core conditions ($\mc$, $\beta$), protostar seed masses ($\mi$) and absorbed fraction of accreting energy ($\alpha$). It is obvious that variations among these parameters will populate the entire HRD regions between the $\sim$1 and $\sim$10 Myr non-accreting isochrones in the low mass star/BD regime.
Work is in progress to perform a more systematic study, assuming a distribution of $\mc$, $\beta$ and $\mi$.
As mentioned above, the collapse of cloud cores with  large $\mc$ and $\beta$ produce the most intense accretion bursts, which can explain the observed properties of  Fu Ori like objects, in particular their SED and the factor $\sim$ 2 increase of the stellar radius. 

\section{Impact on the final mass and IMF determination} \label{IMF}

A natural question arising from the present global scenario is : what is the impact of early accretion upon the mass determination at young ages and thus what is the impact on the IMF determination for young clusters, derived from mass-luminosity relations based on non-accreting isochrones ?

A first interesting feature concerning the impact of accretion upon the final object is that this impact is relatively small in the low-mass ($\simle 0.04 \msun$) BD domain.
First of all, as seen in all figures, the dispersion of luminosity in this domain reduces significantly, as the 1 and 10 Myr non-accreting isochrones tend to merge towards a single line. This stems from the fact that the luminosity barely evolves during this timescale for low-mass BDs with masses $\simle 0.04 \msun$, due to ongoing deuterium burning\footnote{Deuterium burning lasts $\sim$ 9 Myr for a 0.03 $\msol$ BD and $\sim$ 6 Myr for a 0.04 $\msol$ BD according to the Baraffe et al. (1998) models.}. A second (related) feature, as seen in Fig. \ref{fig4}, is that accretion has a lesser impact on the evolution of BDs of final masses $\simle$ 0.04 $\msol$. As explained in Appendix A, the energy released from (initially present and accreted) deuterium  fusion can partly overcome the gravitational energy increase due to mass accretion for these very low masses, 
limiting the total contraction of the object compared to its non-accreting counterpart of same mass and age. 
We expect this to be true as long as deuterium fusion remains efficient, i.e down to  $\sim$ 0.01 $\msun$. Therefore, brown dwarfs in the mass range $\sim 0.01-0.04 \msun$ should be only moderately affected by their accretion histories. Work is under progress to explore the early evolution of accreting BD's below the deuterium burning minimum mass. 


This is different for higher masses.
Table 2 of BCG09 already provides some information about the difference between observed properties, $L$ and $\te$, for accreting and non-accreting objects of given mass in the LMS domain. In order to illustrate the impact of accretion history on the IMF determination, we provide some comparisons in Table 1 and Table 2 between the masses inferred from accreting and non-accreting (Baraffe et al. 1998) models,
respectively, for a few typical examples. Table 1 compares the inferred masses for {\it given luminosity $L$ and effective temperature $\te$} while Table 2 compares the masses for {\it given  age and effective temperature $\te$}.
As seen in Table 1,  inferring the mass for a given ($L$, $\te$) from non-accreting models can yield an {\it overestimation of the mass} by $\sim 20\%$ to 
$\sim 40\%$ for the largest masses. The overestimation can be even larger when inferring the mass from given age and $\te$, as illustrated in Table 2. This is consequential when inferring the IMF for young clusters from the observed signatures in the HRD. Similarly, as seen in the same Table, inferring an age for LMS from their observational signatures  from {\it non-accreting isochrones can lead to ages as large as  20 Myr, whereas the objects are 1 Myr old in reality}. This highlights the crucial impact of early accretion history on the evolutionary properties of young low-mass objects.
As mentioned above, for masses $M \simle 0.04\msun$, the impact is much less severe. A more systematic study exploring a wider range of parameters ($\mc$, $\beta$, $\mi$, $\alpha$) is under progress to further explore this impact on the IMF.

\begin{table*}
\begin{center}
\caption{Comparison between masses  (in $\msun$) inferred from accreting  and non accreting models, respectively,  for some given ($L$, $\te$). The table provides the mass $M_{\rm 1 Myr}$ resulting from accreting sequences which correspond to these ($L$, $\te$) {\it at 1 Myr}, the mass $M_{\rm BCAH98}$ which corresponds to the same ($L$, $\te$) for non accreting models (Baraffe et al. 1998), with an age $t_{\rm BCAH98}$ (in Myr). The last three columns provide the various initial conditions, $\mc$, $\mi$ and $\alpha$, which yield these masses for the accreting sequences.}
\label{table1}
\vskip.1cm
\begin{tabular}{cccccccc}
\hline
$\log L/L_\odot$ & $\te$   & $M_{\rm 1 Myr}$  & $M_{\rm BCAH98}$ & $t_{\rm BCAH98}$ & $\mc$ ($\msun$) & $\mi$ ($\mjup$) & $\alpha$      \\
\hline
\hline
 -2.12 & 2783   &  0.04    & 0.04   & 5     & 0.075 & 1 & 0  \\
-1.919 & 3026  &   0.065 & 0.08   &  10  & 0.085 & 1 & 0  \\
 -1.72 & 3115   &  0.09     &  0.11  & 10     & 0.1 & 1 & 0      \\
 -0.83 & 3750 & 0.5 &  0.7  & 20 & 0.92 & 5 & 0    \\
 -0.74 & 3763    &  0.5       &  0.7 &  15   & 0.92 & 1 & 0        \\
-0.32 & 4182    & 0.9       & 1.1    & 10  & 1.85 & 1 & 0     \\
-0.37 & 4221    & 0.9      &  1.05 & 15     & 1.85 & 5 & 0       \\
\hline
\end{tabular}
\end{center}
\end{table*}

\begin{table*}
\begin{center}
\caption{Same as Table \ref{table1} for a given (age, $\te$). $M_{\rm acc}$ is the mass derived
from an accreting sequence with  $\mc$, $\mi$ and $\alpha$ specified in the table.}
\label{table2}
\vskip.1cm
\begin{tabular}{cccccccc}
\hline
{\rm age (Myr)} & $\te$   & $M_{\rm acc}$  & $M_{\rm BCAH98}$  & $\mc$ ($\msun$) & $\mi$ ($\mjup$) & $\alpha$      \\
\hline
\hline
1 & 2720 & 0.03 & 003 & 0.061 & 1 & 0 \\
1 & 3026 & 0.065 & 0.11 & 0.085 & 1 & 0 \\
1 & 3750 & 0.5 & 0.9 & 0.92 & 5 & 0 \\
1 & 4182 & 0.9 & 1.4 & 1.85 & 1 & 0 \\
\hline
\end{tabular}
\end{center}
\end{table*}

   \section{Conclusion} \label{conclusion}

  In this paper, we have presented a global scenario, based on  
  accretion evolutionary sequences for low-mass stars and brown dwarfs characterized by accretion histories derived from numerical
 hydrodynamics simulations of their parent collapsing cloud prestellar core. This scenario suggests the 
 following paradigm:
 \begin{itemize}
 \item{(i)}  cloud fragmentation and collapse lead to a distribution of collapsing prestellar cores with various masses $\mc$, various ratios of rotational
 to gravitational energy $\beta$, and various magnetic field properties. Each of these collapsing cores
 leads to different episodic accretion histories, characterized by different burst episodes/intensities. As a general feature, the larger $\mc$ and $\beta$, the more prone the disk to fragmentation and the more intense the accretion burst episodes.
 \item{(ii)}  initial protostars (second Larson cores) form at the center of these collapsing prestellar cores when reaching the stellar density ($\sim 10^{-2}$ g cm$^{-3}$) and thus the adiabat determined by H$_2$ dissociation and the  opacity limit for fragmentation. This limit remains presently uncertain but should be of the order of 1 $\mjup$ and should vary only modestly with the cloud conditions, even though the exact amount of variation remains to be determined from appropriate RMHD simulations of the second collapse.
 \item{(iii)} a fairly small or negligible (at most a few \%) amount of accretion energy is expected to be absorbed by the central object 
in the very-low mass (mostly BD) regime, as expected from the moderate accretion rates produced by small-mass collapsing cores, which yield very scarce of even no burst episodes. In contrast, a larger amount of absorbed accretion energy (up to $\sim$ 20\%) is expected for larger initial core masses (thus larger stellar masses)  and large rotation rates. This amount increases with the number of accretion bursts with rates  exceeding some critical value $ \sim 10^{-5}~M_\odot$~yr$^{-1}$. Substantial variations of $\alpha$ in this regime are expected, with $\alpha$ increasing with the core mass and initial rotation rate.  \item{(iv)} in some cases, high-mass and rapidly rotating collapsing cores may lead to such intense bursts that one enters the domain of "hot accretion", with $\alpha \gtrsim 0.2$, yielding a strong increase of the protostar radius and luminosity, typical of Fu Ori like events.
Whether the evolution of $\alpha$ from "cold" to "hot" is determined by a change in shock conditions or by a change of topology in the accretion process remains to be determined and should motivate dedicated studies. 
 \end{itemize}
 
We have shown that this global scenario can explain both (1) the observed luminosity spread in the HRD of young clusters over the entire low-mass ($\le 1\,\msol$) range, with no significant age spread, and (2) the properties (mass, radius) inferred from the spectral energy distribution of Fu Ori objects, providing an explanation to the questions raised by  HZC11. If correct, this unified picture should close the controversy raised by HOK11, except if it can be shown unambiguously that the initial protostar/BD mass can not be smaller than  about 10 $\mjup$, i.e. that the first and second Larson's core are about the same mass, a rather unlikely possibility given the enormous differences in temperatures and densities, thus entropies, during the first and second collapse. 

More work is definitely required to test the above-mentioned paradigm. Regarding assumption (i), the link between the accretion burst intensity and both $\mc$ and $\beta$ requires further studies, ideally with 3D MHD simulations including the effect of magneto-rotational instability in the innermost disk, but as a first step by including self-consistent radiative feedback from the protostar
as discussed in section~\ref{accretion}.  Concerning point (ii), 3D simulations of the second collapse including radiation hydrodynamics and non-ideal MHD are now underway (Machida et al. 2010, Masson et al. 2011). They should provide more insight on the second Larson's core mass determination and characteristic mechanical and thermal properties ({\it e.g} initial entropy content) and on the dependence of these properties upon the cloud initial conditions ({\it e.g} rotation, magnetic field).

Assumptions (iii) and (iv) suggest the existence of a threshold in accretion rates above which a transition from cold to hot accretion should occur. Many possibilities can be speculated to suggest an evolution of the accretion pattern/properties onto the protostar. The simplest explanation, as outlined in Appendix B, is that large accretion rates, orders of magnitude larger than the "standard" $c_S^3/G\sim {10^{-6}\,\msolyr}$ value, are energetic enough (1) to penetrate substantially below the surface layers, within the protostar deeper (convective) layers, (2) to dissipate enough kinetic energy to significantly affect the protostar's heat budget, halting its contraction. Such a transition may also stem from a change in the accretion geometry, evolving for instance from magnetospheric accretion for weak accretion rates, to thick disk accretion for intense bursts, as
suggested by the absence of magnetospheric accretion in Fu Ori (HZC11). It might also stem from a change in the disk properties, evolving from thin disk to thick disk accretion with increasing accretion rates.  
It could also correspond to a transition from a supercritical shock to a (optically thin) subcritical shock at the second core stage (Commer\c con et al. 2011). The topology (and the value) of the protostar's magnetic field may also play a major role in the accretion process; 3D MHD simulations by Long et al. (2008) indeed show that the mass accretion rate from the disk onto the star, the area covered by hot spots and the angular momentum flux from accreting material between the disk and the star are several times smaller in the case of a quadrupole dominated configuration than in the dipole case. The dependence of the protostar magnetic field topology upon its mass could thus have an impact on the value of  $\alpha$.  Finally, increasing the accretion rate yields a smaller magnetospheric radius, at which the disk is truncated by the stars's magnetic field, evolving from accretion through funnels to direct accretion onto the star through a boundary layer (Romanova et al. 2008a). It is indeed known that at high accretion rates characteristic of Fu Ori outburst phases, the magnetospheric radius moves much closer to the stellar surface. High accretion rates are also predicted to yield unstable situations with accretion evolving from funnel streams to a situation with multiple hot spots on the stars's surface (Romanova et al. 2008b, Kulkarni \& Romanova 2008). Exploring these issues requires dedicated observational and theoretical efforts in order to better understand the star-disk interaction and the accretion processes onto protostars. 

The consequences of the present study have also far reaching consequences in the domain of IMF and star formation. First of all, it shows that the concept of a birthline does not apply to low-mass ($\lesssim 1\,\msun$) objects. Each proto-star/BD experiences its own (unpredictable) accretion history and ends up randomly in the HRD at the end of the accretion process. Second of all, inferring masses from observed $L$-$\te$ or $\te$-age properties in the HRD from isochrones which do not take into account the complete
accretion history can yield severely incorrect determinations, possibly overestimating the mass by as much as 40\% or more. The possibility of such errors should be borne in
mind when trying to determine the IMF of young ($\lesssim 10$ Myr) clusters. In the same vein, age determinations of faint objects in clusters or star forming regions from "standard" isochrones can overestimate the age of these objects by several tens of Myr ! As we have shown in BC10 that accretion also strongly increases lithium depletion, this definitely invalidates the arguments in favor of "slow" star formation and should close this debate.

To conclude,  we are conscious that the scenario presented in this paper  (and in BCG09 and BC10) has weaknesses, as it relies on some assumptions which, so far,
lack robust physical fondations.  On the other hand, the scenario relies on 
the treatment of the impact of episodic accretion, {\it inferred} from the collapse of a cloud prestellar core, onto the structure and evolution of the accreting protostar/BD. Within this framework, it provides a consistent explanation for several, apparently uncorrelated, observational features, namely the HRD luminosity and radius (see Jeffries 2007) spread, the properties of Fu Ori, the abnormal Li depletion  of some young objects (see BC10) and the luminosity distribution of embedded protostars (Evans et al. 2009, Dunham \& Vorobyov 2011). This should motivate further observational and theoretical studies to test the reliability of this global picture.

\acknowledgments
The authors are thankful to Lee Hartmann for exciting discussions and to M. Machida for useful e-mail exchanges. This work was supported by the European Research Council under the European CommunityÕs Seventh Framework Programme (FP7/2007-2013 Grant Agreement No. 247060) and by Royal Society awards WM090065 and RFBR Cost shared application with Russia
(JP101297, 10-02-00278 and 11-02-92601).
Numerical simulations were done at the Atlantic Computational Excellence
Network (ACEnet) and the Shared Hierarchical Academic Research Computing
Network (SHARCNET).   EIV acknowledges support from the Lise
Meitner Fellowship.

\appendix
\section{Mass and deuterium accretion}

In this appendix, we explain the results displayed in Fig. \ref{fig1}, namely the increasing effect of accretion on the central object's structure for decreasing $\mi$ at fixed $\mc$, for $\mc=0.085 \msol$.
These results can be understood from arguments suggested by Lee Hartmann, by comparing the binding energy of the central object with the amount of energy available from deuterium fusion (Hartmann, priv. com.).  Assuming $n=3/2$ polytrope, the total binding energy of a protostar is:
\begin{equation}
E_{\rm tot} =- {3 \over 7} {G M^2 \over R}.
\label{eq_tot}
\end{equation}
The fusion reaction of deuterium $p+d$ provides an energy $Q$=5.494 Mev per reaction  (Caughlan \& Fowler 1988). Assuming a mass fraction [$D$] = $2 \times 10^{-5}$, the energy per gram of material available from D burning is thus $\epsilon_{\rm deut}  \sim10^{14}$ erg/g. Assuming complete fusion of D (initially present and accreted), the total energy released from D fusion when the object has reached a mass $M$, is:
\begin{equation}
E_{\rm deut} =  M \times \epsilon_{\rm deut}.
\label{eq_deut}
\end{equation}
If $|\etot |> \edeut$, the release of nuclear energy is insufficient to balance the increase of gravitational energy of the growing protostar, yielding a more compact object than  its non-accreting counterpart of same mass $M$ and age (Stahler 1988; Hartmann et al. 1997). Combining Eqs (\ref{eq_tot}) and  (\ref{eq_deut}), the above-mentioned inequality yields the following condition on the radius of the protostar:
\begin{equation}
R < 8.2 {M \over \msun} R_\odot  \equiv R_{\rm crit}.
\label{eq_R}
\end{equation}
If $R < R_{\rm crit}$, accretion produces an object more compact than the non-accreting counterpart,
{\it provided} $\tmdot < \tau_{\rm KH}$. The latter condition is always fulfilled during burst events (with $\alpha$=0). 
In contrast, if $R > R_{{\rm crit}}$, the energy released by D burning will limit the contraction of the accreting object. 
For the models displayed in Fig. \ref{fig1}, comparison of the models starting with 1 $\mjup$ (hereafter model A) and 5 $\mjup$ (model B), respectively, shows that before the onset of D fusion, model A contracts faster than model B as a result of accretion, for the same accretion rate, on an object with an initially lower binding energy\footnote{Both sequences start with an initial radius $R \sim 1 \rsun$. Variations of the initial radius for values $>$ 0.7 $\rsun$ yield the same conclusions.} and no nuclear energy generation. At the onset of deuterium burning ($T_{\rm c} \sim 6 \times10^5$ K, $t = 10^4$ yr), model A has a larger binding energy  than model B, because of its significantly smaller radius (see Table \ref{tableA}). Deuterium burning and ongoing accretion thus proceed with $R < R_{\rm crit}$ for model A and with $R > R_{\rm crit}$ for model B. Consequently, according to the afore mentioned argument, accretion has a much
larger impact on model A than on model B, yielding  a significantly  more compact structure than the non-accreting counterpart (see Table \ref{tableA}). 

\begin{table}[!ht]
\caption{Properties of accreting models displayed in Fig. \ref{fig1} (see \S \ref{ini}) and starting respectively with 1 $\mjup$ (model A) and 5 $\mjup$ (model B), for $\alpha$=0. Models  with $\mdot=0$ refer to the non-accreting counterparts. Energies are in units of 10$^{45}$ erg. The Kelvin-Helmholtz timescale $\tau_{\rm KH}$ and accretion timescale $\tmdot$  are also provided (for the latter, only for models A and B at 10$^4$ yr when accretion is relevant), in units of Myr.}
\label{tableA}
\begin{tabular}{lcccccccc}
\hline
model & $t$ (yr) & $M$ ($\msun$) & $R$ ($\rsun$)  & $R_{\rm crit}$ ($\rsun$) & $E_{\rm tot}$ & $E_{\rm tdeut}$ & $\tau_{\rm KH}$ (Myr) &  $\tmdot$ (Myr)\\
\hline
A& 10$^4$ & 0.036 & 0.3 & 0.3 & 7.2 & 7.2 & 25 & 0.07\\
B & 10$^4$ & 0.04 & 0.52 & 0.33 & 5.1 & 8.1 & 6 & 0.08 \\
\hline
A & 10$^6$ & 0.065 & 0.4 & 0.54 & 17.5 & 13.1 & 27 & - \\
B & 10$^6$ & 0.07 & 0.65 & 0.57 & 12.2 & 13.9 & 8.5   & - \\
$\mdot =0$ &  10$^6$ & 0.065 & 0.70 &  - & - &  & 6 & -\\
$\mdot =0$ &  10$^6$ & 0.07 & 0.75 & - & - & - & 6 & -\\
\hline
\end{tabular}
\end{table}

The condition (\ref{eq_R}) thus provides a good criterion to infer the effect of accretion during and at the end of the accretion process, for objects massive enough to ignite D-fusion, i.e. $M \simgr 0.013 \msun$ (Chabrier et al. 2000). Note that for model A, at 1 Myr the object still has a radius $R < R_{\rm crit}$, hence looking older than the non-accreting counterpart, whereas for model B the object has a radius $R > R_{\rm crit}$ at this epoch, closer to that of the non-accreting counterpart (see Table \ref{tableA}). 

Criterion  (\ref{eq_R})  explains as well the results displayed in Fig. \ref{fig1bis} (see \S \ref{mc}),  namely the larger the prestellar core mass $\mc$ the more compact the structure of the newly formed object compared to its non-accreting counterpart at 1 Myr (see Table \ref{tableB}). Indeed, for a given (same) initial protostar mass $\mi$, larger prestellar core masses $\mc$ yield objects with a larger final mass and thus a larger binding energy, so that criterion 
(\ref{eq_R}) is more easily fulfilled (see Table \ref{tableB}). As mentioned above, these arguments hold as long as the forming object is massive enough to fuse deuterium. 

\begin{table}[!ht]
\caption{Properties at 1 Myr of accreting models of Fig. \ref{fig1bis} (see \S \ref{mc}), starting with different initial prestellar core masses $\mc$ and same $\mi=1 \mjup$, and of non-accreting counterparts.}
\label{tableB}
\begin{tabular}{lcccccc}
\hline
model & $M$ ($\msun$) & $R$ ($\rsun$)  & $R_{\rm crit}$ ($\rsun$) & $E_{\rm tot}$ (10$^{45}$ erg) & $E_{\rm tdeut}$ (10$^{45}$ erg)\\
\hline
$\mc = 0.061 \msun$ &  0.03 & 0.35 & 0.26 & 4.84 & 6.46 \\
 $\mc = 0.075 \msun$ &  0.04 & 0.38 & 0.33 & 7.1 & 8.1 \\
 $\mc = 0.085 \msun$ &  0.065 & 0.4 & 0.54 & 17.5 & 13.1 \\
 $\mc = 0.1 \msun$ &  0.09 & 0.48 & 0.74 & 28 & 18 \\
$\mdot =0$ &  0.03 & 0.49 & - & - & \\
 $\mdot =0$ &  0.04 & 0.5 & - & - & \\
$\mdot =0$ &   0.065 & 0.7 & - & - & \\
$\mdot =0$ &   0.09 & 0.91 & - & - & \\
\hline
\end{tabular}
\end{table}

\section{Properties at the shock front}

In this appendix, we derive analytical estimates for the various properties of the flow at the shock front. 
All quantities are expressed in cgs units.

$\bullet$ {\it Pressure of the accretion flow}

The ram pressure of the accretion flow, falling down on the proto-star/BD at the free-fall/Keplerian velocity $v_{\rm acc}=(2GM/R)^{1/2}$, can be estimated as :
\begin{equation}
P_{\rm acc} 
= \frac{\mdot \, v_{\rm acc}} {8\pi R^2  \delta} \approx \frac{ 3\times 10^{4}} {\delta} (\frac{\mdot}{10^{-6}\,\msolyr})(\frac{M}{\msol})^{1/2} (\frac{R}{\rsun})^{-5/2}
\label{eq_pram}
\end{equation}
Assuming a stellar surface fraction $\delta\approx 0.1$ covered by accretion, this yields $P_{\rm acc}\approx 10^4\,(\frac{\mdot}{10^{-6}\,\msolyr})$ dyne cm$^{-2}$ for a 1 to 10 $\mjup$ initial protostar.

$\bullet$ {\it Properties of the photosphere}

For an atmosphere in hydrostatic equilibrium, $\nabla P=-\rho g$, where $g=2.7\times 10^4(\frac{M}{\msol})(\frac{R}{\rsun})^{-2}$ is the surface gravity, the optical depth at depth $z$, $\tau(z)=z/l_\nu$, where $l_\nu=1/(\rho{\bar \kappa})$ is the photon mean free path and ${\bar \kappa}$ the Rosseland mean opacity, obeys the condition
\begin{equation}
\tau(z)=\int_z^\infty \frac{{\bar \kappa}}{g}dP.
\label{eq_tau}
\end{equation}
The pressure and density at the photosphere, $\tau\sim 1$, are thus given by
\begin{eqnarray}
P_{\rm ph} &\approx& \frac{g}{{\bar \kappa}}\approx 10^4 \,(\frac{g}{10^2})(\frac {{\bar \kappa}}{10^{-2}})^{-1}\label{eq_pphot} \\
 \rho_{\rm ph} &=&\frac{\mu}{{\cal R}}\frac {P_{\rm ph}}{\te} \approx 10^{-7} \mu\,  (\frac{P}{10^4})(\frac{\te}{10^3\,{\rm K}})^{-1}, 
 \label{eq_rphot}
\end{eqnarray}
where $\mu$ denotes the mean molecular weight and ${\cal R}$ the perfect gas constant.
\noindent At $\sim 10^{-6}\,\msolyr$, the 1$\mjup$ proto-star/BD core quickly reaches a few $\mjup$, which yields $\te \gtrsim 2000$ K and, for an initial radius $\sim 1\rsun$, $\log g\sim 2.0$-2.5 from $\sim$a few to about 10 $\mjup$. 
In this regime, for $R=\rho/T_6^3\sim 10$-100, where $T_6$ is the temperature in units of $10^6$ K, the Rosseland mean opacity is dominated by molecular lines and is
${\bar \kappa}\approx 10^{-2}$ cm$^2$g$^{-1}$ 
(Fergusson et al. 2005). This yields $P_{\rm ph}\approx 10^4$ dyne cm$^{-2}$ for the aforementioned mass range, in good agreement with the values obtained with the complete calculations (see Fig. 4 of Baraffe et al. 2002).

$\bullet$ {\it Low accreting rates}

Let first consider small collapsing prestellar cores ($\mc \lesssim 0.1\,\msun$). Assuming a $\sim$30\%-70\% core-star mass conversion efficiency, these cores will produce final objects in the BD domain. As mentioned in the text, these collapsing cores do not produce high accretion bursts and the accretion rate remains essentially of the order 
of ${10^{-6}\,\msolyr}$ (see e.g. inset in Fig. 1).
According to the above estimates, the pressure at the accretion shock thus remains of the order of or at most about an order of magnitude larger than the photospheric pressure, as the object builds up its mass, from $\sim$ a few $\mjup$ to, eventually, a few tens of $\mjup$. As will be shown below (see eqn.(\ref{eq_delR})), for such conditions, the shock 
 is likely to remain nearly superficial and thus to entail a negligible amount of energy compared with the proto-star/BD binding energy. This leads support to a "cold accretion" condition, with a value $\alpha \ll 1$ for the fraction of absorbed accreting energy in the very low mass domain.

$\bullet$ {\it High accreting bursts}

As the mass of the collapsing cores, and thus of the final objects, increases, the simulations predict an increasing number of bursts, many of them reaching  the $10^{-5}$-$10^{-4}\,\msolyr$ level, or even $10^{-3}\,\msolyr$ for the most intense bursts (see inset in Fig. 3). For such rates, a 10 $\mjup$ accreting protostar mass is built up within $\sim 10$ to 10$^3$ yr. It is easily seen from eqns.(\ref{eq_pram}) and (\ref{eq_pphot}) that in that case $P_{\rm acc}\approx$(10-1000)$\times P_{\rm ph}$, i.e. $\approx 10^5$-10$^7$ dyne cm$^{-2}$, the upper values corresponding to the most intense burst events. 
Eqn.(\ref{eq_tau}) predicts that, {\it at constant mean opacity}, $\tau \propto P$, i.e. a factor $\sim 10$-1000 increase in pressure corresponds to about the same increase in optical depth. As temperature increases with pressure, however, the Rosseland opacity increases (for $R\sim 10$-100) by a factor $\sim 10$ between $\sim$ 2000 and $\sim 4000$ K (Fergusson et al. 2005). The optical depth at the shock front is thus likely to be of the order of $\tau_{\rm acc}\approx 10^2-10^4$. Therefore, even though the shock energy, if the shock is supercritical, will be entirely radiated away {\it at the shock front}, this energy will be at least partly absorbed by the radiative precursor, since the shock occurs in a very optically-thick medium. At these levels, the thermal gradient is adiabatic, implying that transport by convective motions is more efficient than by radiative diffusion, so that the dissipated shock energy will eventually contribute to the proto-star/BD internal heat budget. We can try to estimate this amount of absorbed energy. 

In the optically-thick regime, the temperature at the shock front, $T_{\rm acc}$, can be crudely estimated as: $T_{\rm acc}\approx \te\,\tau_{\rm acc}^{1/4}$ (e.g. Hartmann et al. 1997), which yields $T_{\rm acc}\approx 3500$-10000 K for the aforementioned $P_{\rm acc}/P_{\rm ph}$ range. The density at the shock front is thus: $\rho_{\rm acc}\approx \frac{\mu}{{\cal R}}\frac {P_{\rm acc}}{T_{\rm acc}} \approx 10^{-6}$-$10^{-5}$ g cm$^{-1}$. 

Using the hydrostatic equilibrium equation, $\nabla P=\rho \nabla V(r)$, where $V(r)=GM/r$ is the gravitational potential at distance $r$ from the center, and assuming $\Delta R=R_\star-R_{\rm acc}\ll R_\star$, where $R_\star$ denotes the protostar radius and $R_{\rm acc}$ the radius at the shock front, so that $\Delta R$ represents the depth penetration of the shock, one gets the following rough estimate:
\begin{eqnarray}
\frac {\Delta R}{R} \approx \frac{R_\star \Delta P}{\rho GM} \approx 5\times 10^{-4} \times (\frac{P_{\rm acc}}{10^6}) (\frac {\rho_{\rm acc}}{10^{-6}})^{-1}
 (\frac{M}{\msol})^{-1} (\frac{R}{\rsun})
 \label{eq_delR}
\end{eqnarray}
where $\Delta P= P_{\rm acc}-P_{\rm ph}\approx P_{\rm acc}$ (see above), and $\rho \approx \rho_{\rm acc}$. For $P_{\rm acc}\approx 10^6$ dyne cm$^{-2}$ (see above) this yields $\Delta R/R\sim 0.05$, while for the most intense accretion bursts, with $ P_{\rm acc}\approx 10^6-10^7$ dyne cm$^{-2}$, 
one gets $   {\Delta R}/{R} \simgr 0.1$, for $\sim 10\,\mjup$. 
According to these estimates, the shock, for such high accretion bursts can thus penetrate deep enough below the stellar surface to reach the internal adiabat, which almost reaches the photosphere for low-mass objects. As mentioned above, the absorbed energy should thus be efficiently redistributed
within the entire protostar interior, modifying its energy budget. The corresponding fraction of accreting energy absorbed by the protostar, compared to its binding energy $E$, can be estimated as

\begin{eqnarray}
\frac{U_{\rm acc}}{E} &\approx&  \frac{{\tilde c_V} T_{\rm acc}} {{\tilde c_V} \langle T\rangle} \approx  \frac {(3-10)\,\te}{\langle T\rangle} \approx 0.03-0.10,
 \label{eq_delU}
\end{eqnarray}
where ${\tilde c_V}$ is the specific heat per unit mass and $\langle T\rangle\approx \frac {|E|}{M\,{\tilde c_V}}\approx  10^7(\frac{M}{\msol})(\frac{R}{\rsun})^{-1}
(\frac{{\tilde c_V}}{10^8})^{-1}$ K. Larger accretion rates, as expected for the largest cores, will lead to even larger energy contributions. These estimates tend to corroborate  corresponding values of $\alpha$ for masses issued from high burst accretion phases, as indeed suggested in the text for the Fu Ori events.

As mentioned above, these estimates should be taken with great caution and detailed numerical calculations of the accretion shock properties at the second core stage are strongly needed in order to get more accurate determinations. These estimates, however, seem to bring support to the results obtained in the calculations presented in the text, namely: 

\indent (i) a $\alpha \rightarrow 0$ value for the lowest mass objects (mainly within the BD domain), as these objects result from the collapse of prestellar cores which hardly ever produce accretion rates exceeding the mean value $c_S^3/G\sim {10^{-6}\,\msolyr}$. This leads to accretion shocks which remain located near the photosphere level or at worse dissipate a negligible amount of energy within the protostar interior.

\indent (ii) possible values $\alpha \gtrsim 0.1$ for higher mass cores, thus higher mass stars, within the stellar domain, since the collapse of these cores can lead to accretion bursts several orders of magnitude larger than the aforementioned mean value. Such shocks can penetrate within a substantial fraction of the forming protostar and thus dissipate their energy in the inner regions, contributing non negligibly  to the protostar's energy budget, halting or even counteracting its contraction.

 This globally suggests cold accretion conditions ($\alpha \rightarrow 0$) for the very low mass domain and possible "hot accretion" events for higher masses.

\end{document}